\documentstyle[11pt]{article}

\def\kon#1#2{\vbox{\halign{##&&##\cr\lower4pt
\hbox{$\scriptscriptstyle\vert$}\hrulefill &\hrulefill\lower4pt
\hbox{$\scriptscriptstyle\vert$}\cr $#1$&$#2$\cr}}}

\def\fii{\varphi}
\def\al{\alpha}
\def\be{\beta}
\def\ga{\gamma}
\def\ro{\varrho}

\def\eh{{\scriptstyle{1\over 2}}}
\def\d{\partial}
\def\=d{\,{\buildrel\rm def\over =}\,}

\def\sqr#1#2{{\vcenter{\vbox{\hrule height.#2pt\hbox{\vrule width.
#2pt height#1pt \kern#1pt \vrule width.#2pt}\hrule height.#2pt}}}}

\def\la{\lambda}

\def\sq{\hbox{\rlap{$\sqcap$}$\sqcup$}}

\def\te{\vartheta}

\begin{document}

\title{Dark matter in galaxies according to the tensor-four-scalars theory }
\author{G\"unter Scharf
\footnote{e-mail: scharf@physik.unizh.ch}
\\ Institut f\"ur Theoretische Physik, 
\\ Universit\"at Z\"urich, 
\\ Winterthurerstr. 190 , CH-8057 Z\"urich, Switzerland}

\date{}

\maketitle\vskip 3cm

\begin{abstract} Massive gravity previously constructed as the
spin-2 quantum gauge theory leads in the mass zero limit
to a modification of general relativity. As a relic from the
massive theory a vector field $v^\la$ survives which couples to the
metric only. However, the coupling does not involve covariant
derivatives so that $v^\la$ must be considered as 4 scalar fields. 
We analyze the static, spherically symmetric solutions of this
theory. From the corresponding geodesics we find the circular
velocity profile. Interpreting this as coming from a dark density
profile, the theory predicts a flat density core for $r\to 0$.
But the dark density profile is not universal.

\end{abstract}

\newpage

\section{Introduction}

For some years we have studied massive gravity on the basis
of a cohomological formulation of
gauge invariance in terms of asymptotic free fields. This method works
for massless and massive gauge theories equally well [1], the Higgs
mechanism is not used.
The massive spin-2 theory was first investigated
in [2]. The most elegant way to obtain the theory is by assuming the
gauge invariance condition for all chronological products in the form of
the descent equations [3]. These give the total interaction Lagrangian
including ghost couplings and the necessary coupling to a vector-graviton
field $v^\la$. This vector-graviton field is of fundamental
importance because without it a massive spin-2 gauge theory is
impossible. So the main purpose of this paper is to analyze the
physical consequences of this new actor in the drama of gravity 

Einstein's theory describes reality very well on the scale of the solar system,
that means on distances up to 100 AU. On the scale of galaxies, i.e. kilo-parsec,
one observes deviations from general relativity which are usually ascribed to
hypothetical dark matter. In the tensor-four-scalars theory they come from the
vector-graviton field $v^\lambda$ which must be considered as four scalar fields
in the classical theory. Since the scalar fields $v^\lambda$ and gravity are
coupled by derivative couplings it is the spatial variation of $v^\lambda$ which
matters. From the two length scales above it is plausible that the mean spatial
variation of $v^\lambda$ over a galaxy is $10^6$ times bigger than in the solar system.
This explains why the deviations from general relativity are only observable on the
galactic scale or bigger.

The paper is organized as follows. In the next section we recall the origin
of the modified gravity theory. In section 3 we introduce a galaxy model
and construct the static, spherically symmetric (vacuum) solution. The latter
was already tried in a previous paper [4] with an ansatz containing two free
metric functions. This was not good enough, we need three functions.
In section 4 we compute the power series expansions of the solution 
for $r\to 0$ and $r\to\infty$. In section 5 we consider the geodesics corresponding to
our metric and specialize to circular motion. This gives an expression of the
circular velocity in terms of the metric functions. Interpreting this as
the velocity curve of dark ``matter'' we find a flat density profile for $r\to 0$.
In section 6 we discuss the degeneracy in the field equations. As a consequence
we find that the dark density profile is not universal.

\section{Massive quantum gravity and the tensor-four-scalars theory}

Our starting point is the massive spin-2 quantum gauge theory on Minkowski
space which we have called massive gravity for short.
The basic free asymptotic fields are the symmetric
tensor field $h^{\mu\nu}(x)$ with arbitrary trace, the fermionic ghost
$u^\mu(x)$ and anti-ghost $\tilde u^\mu(x)$ fields and the
vector-graviton field $v^\lambda(x)$. They all satisfy the Klein-Gordon 
equation
$$(\sq+m^2)h^{\mu\nu}=0=(\sq+m^2)u^\mu=(\sq+m^2)\tilde u^\mu
=(\sq+m^2)v^\la    \eqno(2.1)$$
and are quantized as follows [2] [3]
$$[h^{\alpha\beta}(x), h^{\mu\nu}(y)]=-{i\over 2}
(\eta^{\alpha\mu}\eta^{\beta\nu}+\eta^{\alpha\nu}
\eta^{\beta\mu}-\eta^{\alpha\beta}\eta^{\mu\nu})D_m(x-y)
\eqno(2.2)$$
$$\{u^\mu(x),\tilde u^\nu(y)\}=i\eta^{\mu\nu}D_m(x-y)$$
$$[v^\mu(x),\, v^\nu(y)]={i\over 2}\eta^{\mu\nu}D_{m}(x-y),$$ 
and zero otherwise. Here, $D_m(x)$ is the Jordan-Pauli distribution 
with mass $m$ and $\eta^{\mu\nu}={\rm diag}(1,-1,-1,-1)$ the Minkowski 
tensor 

The gauge structure on these fields is defined through a nilpotent
gauge charge operator $Q$ satisfying
$$Q^2=0, \quad Q\Omega=0\eqno(2.3)$$
where $\Omega$ is the Fock vacuum and
$$d_Q h^{\mu\nu}= [Q,h^{\mu\nu}]=-{i\over 2}(\d^\nu u^\mu+\d^\mu u^\nu
-\eta^{\mu\nu}\d_\al u^\al) \eqno(2.4)$$
$$d_Q u^\mu\=d \{Q,u\}=0$$
$$d_Q\tilde u^\mu\=d \{Q,\tilde u^\mu\} =i(\d_\nu h^{\mu\nu}-m v^\mu)
\eqno(2.5)$$
$$d_Q v^\mu\=d [Q, v^\mu]=-{i\over 2}mu^\mu.\eqno(2.6)$$
The vector-graviton field $v^\la$ is necessary for nilpotency of $Q$.
In fact, in order to get $d^2_Q\tilde u=0$ from (2.5), the additional
term $-mv^\mu$ is indispensable. 

The coupling $T(x)$ between these fields follows from the gauge
invariance condition [2]
$$d_Q T(x)=i\d_\al T^\al(x)\eqno(2.7)$$
where $T$ and $T^\al$ are normally ordered polynomials with ghost
number 0 and 1, respectively. In addition we may require the descent
equations 
$$\d_QT^\al=[Q,T^\al]=i\d_\be T^{\al\be}\eqno(2.8)$$
$$[Q,T^{\al\be}]=i\d_\ga T^{\al\be\ga}\eqno(2.9)$$
where the new $T$'s are antisymmetric in the Lorentz indices.
The essentially unique coupling derived from (2.7-9) is given by [3]
$$T=h^{\al\be}\d_\al h\d_\be h-2h^{\al\be}\d_\al h^{\mu\nu}\d_\be 
h_{\mu\nu}-4h_{\al\be}\d_\nu h^{\be\mu}\d_\mu h^{\al\nu}$$
$$-2h^{\al\be}\d_\mu h_{\al\be}\d^\mu h+4h_{\al\be}\d^\nu h^{\al\mu}
\d_\nu h_\mu^\be+4h^{\al\be}\d_\al v_\la\d_\be v^\la$$
$$+4u^\mu\d_\be\tilde u_\nu\d_\mu h^{\nu\be}-4\d_\nu u_\be\d_\mu\tilde u^\be 
h^{\mu\nu}+4\d_\nu u^\nu\d_\mu\tilde u_\be h^{\be\mu}$$
$$-4\d_\nu u^\mu\d_\mu\tilde u_\be h^{\nu\be} 
-4mu^\al\tilde u^\be\d_\al v_\be-m^2\Bigl({4\over 3}h_{\mu\nu} 
h^{\mu\be}h^\nu_\be$$
$$-h^{\mu\be}h_{\mu\be}h+{1\over 6}h^3\Bigl).\eqno(2.10)$$
Here $h=h_\mu^\mu$ is the trace and a coupling constant is arbitrary.
The quartic couplings follow from second order gauge invariance and so on.

We consider the limit $m\to 0$ in the following. The massless limit of
massive gravity is certainly a possible alternative to general relativity.
The new physics comes from the surviving coupling term of the
vector-graviton
$$T_v=4h^{\al\be}\d_\al v_\la\d_\be v^\la.\eqno(2.11)$$
To be able to do non-perturbative calculations we look for the classical
theory corresponding to the coupling (2.10). It was shown in [1]
sect.5.5 that the pure graviton couplings $h\d h\d h$ correspond to the
Einstein-Hilbert Lagrangian
$$L_{\rm EH}=-{2\over\kappa^2}\sqrt{-g}R$$
in the following sense. We write the metric tensor as
$$\sqrt{-g}g^{\mu\nu}=\eta^{\mu\nu}+\kappa h^{\mu\nu}$$
and expand $L_{\rm EH}$ in powers of $\kappa$. Then the quadratic terms
$O(\kappa^0)$ give the free theory, the cubic terms $O(\kappa^1)$ agree
with the pure graviton coupling terms in (2.10) up to a factor and 
and quartic and higher couplings follow from higher order gauge invariance,
so that the full non-linear structure of general relativity is recovered.

To obtain $T_v$ (2.11) we must add
$$L_v=-\sqrt{-g}g^{\al\be}\d_\al v_\la\d_\be v^\la\eqno(2.12)$$
to $L_{\rm EH}$. One may be tempted to write covariant derivatives
$\nabla_\al$ instead of partial derivatives in order to get a true
scalar under general coordinate transformations. But this would produce
quartic couplings containing $v_\la$ and such terms are absent in the
quantum theory ([2], eq.(4.12)). For the same reason the Lorentz index
$\la$ in $v_\la$ is raised and lowered with the Minkowski tensor
$\eta_{\mu\nu}$, but all other indices with $g_{\al\be}$. {\it Both together
means that the vector graviton field $v_\lambda$ should be considered
as four scalar fields in the classical theory.} Then (2.12) is a scalar
under general coordinate transformations as it must be.

Of course we want to include ordinary matter in the theory which we describe 
by a complex scalar field $\fii$ 
of mass $M$. In the Klein-Gordon equation for this field we retain Planck's constant
$\hbar$ different from 1. In the end we want consider the classical limit $\hbar\to 0$. Then
in the semi-classical approximation the solution of the Klein-Gordon equation is of the form
$$\fii=\sqrt{\ro}\exp\Bigl({i\over \hbar}S\Bigl)\eqno(2.13)$$
and it has the following interpretation: $\ro(x)$ is the density of particles in space
and time and
$$\vec p={\d S\over \d \vec x}\eqno(2.14)$$
is the momentum of the particle at $x$. $S(x)$ is the Hamilton-Jacobi principal function.
Nobody forbids to take for $M$ the solar mass so that the particles are the stars in a
galaxy, In this way we get a simple model of a galaxy. It is indeed simple because all
stars have the same mass as the sun and gas etc. is neglected. On the other hand
the coupling of the complex scalar field to gravity is completely known from the analysis
of gauge invariance [1]; we do not need equations of state as in more
realistic models. The total Lagrangian of our model now reads
$$
L_{\rm tot}={-2\over\kappa^2}\sqrt{-g}R+{1\over 4}\sqrt{-g}\Bigl[g^{\mu\nu}\hbar^2
(\d_\mu\fii^+\d_\nu\fii+\d_\mu\fii\d_\nu\fii^+)-2M^2\fii^+\fii\Bigl]-$$
$$-\sqrt{-g}g^{\mu\nu}\d_\mu v_\lambda\d_\nu v^\lambda.\eqno(2.15)$$

The Euler-Lagrange equations for the Lagrangian (2.15) give the system of coupled
field equations. Variation of 
$
g^{\mu\nu}
$ 
yields the modified Einstein equations
$$R_{\mu\nu}-{1\over 2}g_{\mu\nu}R=$$
$$={16\pi G\over c^3}\Bigl\{{\hbar^2\over 4}(\d_\mu\fii^+
\d_\nu\fii+\d_\mu\fii\d_\nu\fii^+)-{1\over 4}g_{\mu\nu}(\hbar^2g^{\al\beta}\d_\al\fii^+
\d_\beta\fii-M^2\fii^+\fii)-$$
$$-\d_\mu v_\lambda\d_\nu v^\lambda
+{1\over 2}g_{\mu\nu}g^{\al\beta}\d_\al v_\lambda\d_\beta v^\lambda\Bigl\}\eqno(2.16)$$
The variational derivative with respect to $v^\mu$ gives the wave equation in the metric 
$g^{\al\beta}$
$$
2\d_\al(\sqrt{-g}g^{\al\beta}\d_\beta v_\mu)=0.\eqno(2.17)$$
Finally, the variation of 
$\fii^+$ where $\fii$ is not varied 
gives the Klein-Gordon equation in the metric 
$
g^{\al\beta}
$ 
$$
{\hbar^2\over\sqrt{-g}}\d_\al(\sqrt{-g}g^{\al\beta}\d_\beta \fii)+M^2\fii= 0.
\eqno(2.18)$$

From the point of view of quantum gauge invariance this rather simple tensor-four-scalars 
theory has the same right for being considered as fundamental as Einstein's theory. The
latter is the somewhat exceptional $m=0$ theory; the former is the massless limit of the
massive spin-2 gauge theory. 

\section{Static spherically symmetric solutions}

As a first step to analyze the tensor-four-scalars theory we study vacuum solutions
of the field equations (2.16) (2.17) neglecting the normal matter ($\fii=0$). Of course
the Schwarzschild solution of ordinary general relativity is also a solution of our
theory with $v^\lambda=$ const. This solution describes reality very well in the solar system.
But on larger scales the spatial variation of $v^\la$ becomes observable, so that the
source terms depending on derivatives of $v^\la$ must be taken into account. In this sense,
{\it Einstein's theory can be considered as the local version of the tensor-four-scalars theory.}
It is our aim now to study solutions with non-constant $v^\la$. We choose
spherical coordinates
$$x^0=ct,\quad x^1=r,\quad x^2=\te,\quad x^3=\phi\eqno(3.1)$$
and assume the metric $ds^2=g_{\mu\nu}dx^\mu dx^\nu$ to be of the following static
spherically symmetric form
$$ds^2=e^ac^2dt^2-e^bdr^2-r^2e^c(d\te^2+\sin^2\te d\phi^2),\eqno(3.2)$$
where $a(r), b(r), c(r)$ are functions of $r$ only. Then the determinant of the metric 
is equal to
$$g=-e^dr^4\sin^2\te,\quad d=a+b+2c\eqno(3.3)$$
and we obtain the following non-vanishing Christoffel symbols: 
$$\Gamma^0_{10}={a'\over 2},\quad \Gamma^1_{00}={a'\over 2}e^{a-b}\eqno(3.4)$$
$$\Gamma^1_{11}={b'\over 2},\quad \Gamma^1_{22}=-(r+{r^2\over 2}c')e^{c-b}\eqno(3.5)$$
$$\Gamma^1_{33}=\sin^2\te\Gamma^1_{22},\quad \Gamma^2_{12}={1\over r}+{c'\over 2}=
\Gamma^3_{13}\eqno(3.6)$$
$$\Gamma^2_{33}=-\sin\te\cos\te,\quad \Gamma^3_{23}=\cot\te.\eqno(3.7)$$
The prime always means partial derivative with respect to $r$.

The next step is the computation of the Ricci tensor. Only the diagonal elements
are different from zero:
$$R_{00}=\eh e^{a-b}(a''+\eh a'^2-\eh a'b'+a'c'+{2\over r}a')\eqno(3.8)$$
$$R_{11}=-\eh(a''+2c'')+{b'\over 4}(a'+2c'+{4\over r})-{a'^2\over 4}-{c'^2\over 2}
-{2\over r}c'\eqno(3.9)$$
$$R_{22}=e^{c-b}[-1-{r^2\over 2}c''-r(2c'+{a'-b'\over2})-{r^2\over4}c'(a'-b'+2c')]+1
\eqno(3.10)$$
$$R_{33}=\sin^2\te R_{22}.\eqno(3.11)$$
This gives the following scalar curvature
$$R=g^{\mu\nu}R_{\mu\nu}$$
$$=e^{-b}[a''+2c''+{a'^2\over 2}-{a'b'\over 2}+a'c'+{2\over r}a'-{2\over r}b'-b'c'
+{3\over 2}c'^2+{6\over r}c']+{2\over r}(e^{-b}-e^{-c}).\eqno(3.12)$$
Now we are ready to calculate the left-hand side of the modified Einstein equation (2.16)
$$G_{\mu\nu}=R_{\mu\nu}-\eh g_{\mu\nu}R.\eqno(3.13)$$
It is convenient to raise one index by multiplication with $g^{\mu\nu}$:
$$G_0^0=e^{-b}(-c''-{3\over 4}c'^2+{b'c'\over 2}+{b'-3c'\over r}-{1\over r^2})
+{e^{-c}\over r^2}\eqno(3.14)$$
$$G_1^1=e^{-b}(-{a'c'\over 2}-{a'+c'\over r}-{c'^2\over 4}-{1\over r^2})+{e^{-c}\over r^2}
\eqno(3.15)$$
$$G_2^2=e^{-b}(-{a''\over 2}-{c''\over 2}-{a'^2\over 4}+{a'b'\over 4}-{a'c'\over 4}
+{b'c'\over 4}-{c'^2\over 4}+{b'-a'-2c'\over 2r})=G_3^3.\eqno(3.16)$$

Before we calculate the right-hand side of the modified Einstein's equation we consider
the wave operator in the metric (2.17) neglecting the time derivatives $\d_0$:
$$\d_\al (\sqrt{-g}g^{\al\beta}\d_\beta))=-\sin\te\d_r\Bigl(e^{d/2-b}r^2\d_r\Bigl)
-e^{d/2-c}\d_\te(\sin\te\d_\te)
-{e^{d/2-c}\over\sin\te}\d^2_\phi.\eqno(3.17)$$
Then the field equation for a static $v^\lambda$ becomes
$${\d\over\d r}\Bigl(e^{d/2-b}r^2{\d\over\d r}\Bigl)v^\lambda=e^{d/2-c}L^2v^\lambda,
\eqno(3.18)$$
where $L^2$ is the quantum mechanical angular momentum operator squared. Since we 
restrict ourselves to spherically symmetric solutions the right-hand side vanishes.
Then only the time component $v^\lambda=(v_0(r),0,0,0)$ can be different from 0,
otherwise we get non-diagonal elements in the metric.
The equation (3.18) can now be integrated once
$$v'_0(r)={A\over r^2}e^{b-d/2},\eqno(3.19)$$
where $A$ is an integration constant. With this result the modified Einstein equations
(2.16) assume the following form:
$$G_0^{\,0}=-{\tilde g\over 2}e^{-b}{A^2\over r^4}e^{2b-d}\eqno(3.20)$$
$$G_1^{\,1}={\tilde g\over 2}{A^2\over r^4}e^{b-d}\eqno(3.21)$$
$$G_2^{\,2}=-{\tilde g\over 2}{A^2\over r^4}e^{b-d}\eqno(3.22)$$
where
$$\tilde g={16\pi G\over c^3}\eqno(3.23)$$
is essentially Newton's constant. Using the above results for the Einstein tensor 
$G_\mu^{\,\nu}$ and multiplying by $e^b$ we get the following three differential equations  
$$c''=-{3\over 4}c'^2+{1\over 2}b'c'+{1\over r}(b'-3c')+{1\over r^2}(e^{b-c}-1)
+{\al\over 2r^4}e^{2b-d}\eqno(3.24)$$
$$0={1\over 2}a'c'+{1\over r}(a'+c')+{c'^2\over 4}+{1\over r^2}\Bigl(1-e^{b-c}\Bigl)
+{\al\over 2r^4}e^{2b-d}\eqno(3.25)$$
$$a''+c''={1\over r}(b'-a'-2c')-{1\over 2}(a'^2-a'b'+a'c'-b'c'+c'^2)
+{\al\over r^4}e^{2b-d},\eqno(3.26)$$
where
$$\al=\tilde gA^2\eqno(3.27)$$
is another form of the constant of integration for $v_0(r)$.

\section{Power series expansions}

At first sight the three equations (4.24-26) for the three functions $a(r), b(r)$ and $c(r)$
seem to look quite awful because of the $r^{-4}$-singularity which comes from the $v_0$-function.
To study the solution for small $r$ we try to solve the equations by power series
$$a(r)=\sum_{n=0}^\infty a_n r^n\eqno4.1)$$
$$b(r)=\sum_{n=0}^\infty b_n r^n.\eqno(4.2)$$
However in $c(r)$ we need a logarithmic term
$$c(r)=-2\log {r\over r_c}+\sum_{n=0}^\infty c_n r^n\eqno(4.3)$$
in order to cancel the singularities. We expand everything in powers of $r$, for example
$$e^{b-c}=\Bigl({r\over r_c}\Bigl)^2e^{b_0-c_0}\Bigl[1+(b_1-c_1)r+r^2\Bigl(b_2-c_2
+{b_1^2+c_1^2\over 2}-b_1c_1\Bigl)\Bigl]+\ldots\eqno(4.4)$$
$$e^{b-a-2c}=\Bigl({r\over r_c}\Bigl)^4e^{b_0-a_0-2c_0}\Bigl[1+(b_1-a_1-2c_1)r+$$
$$+r^2\Bigl(b_2-a_2
-2c_2+{a_1^2+b_1^2+4c_1^2\over 2}-a_1b_1-2b_1c_1+2a_1c_1\Bigl)\Bigl]+\ldots\eqno(4.5)$$
For zeroth order we introduce the two parameters
$$\beta={1\over r_c^2}e^{b_0-c_0}\eqno(4.6)$$
$$\gamma={\al\over r_c^4}e^{b_0-a_0-2c_0}.\eqno(4.7)$$

It is convenient to subtract (3.24) from (3.26)
$$a''={c'^2\over 4}-{a'\over 2}(a'-b'+c')+{1\over r}(c'-a')+{1\over r^2}(1-e^{b-c})$$
$$+{\al\over 2r^4}e^{2b-d}\eqno(4.8)$$
and to use this equation instead of (3.26). Then to lowest order $O(r^{-1})$ all three
equations (6.2.24), (6.2.25) and (4.8) are identically satisfied without giving a
restriction on $a_1, b_1, c_1$. In next order $O(r^0)$ we obtain
$$a_1=-{c_1\over 2}+{1\over c_1}(2\beta-\gamma)\eqno(4.9)$$
$$a_2={c_1^2\over 8}-{a_1^2\over 4}+{a_1b_1\over 4}-{a_1c_1\over 4}-{\beta\over 2}
+{\gamma\over 4}$$
$$={3\over 16}c_1^2-{b_1c_1\over 8}+(2\beta-\gamma){b_1\over 4c_1}-{(2\beta-\gamma)^2
\over 4c_1^2}-{\beta\over 2}+{\gamma\over 4}\eqno(4.10)$$
$$c_2=-{3\over 8}c_1^2+{b_1c_1\over 4}+{\beta\over 2}+{\gamma\over 4}.\eqno(4.11)$$
In the order $O(r)$ we find
$$a_3=-{3\over 32}(c_1^3-c_1^2b_1)-{c_1^2b_1\over 48}-{7\over 24}\beta c_1-{\gamma\over 16}
c_1+{\beta\over 12}{b_1^2\over c_1}-{\gamma\over 24}{b_1^2\over c_1}+{\beta^2\over 6c_1}
-{\beta\gamma\over 2c_1}+$$
$$+{5\over 24}{\gamma^2\over c_1}-{\beta^2\over 2}{b_1\over c_1^2}+{\beta\gamma\over 2}
{b_1\over c_1^2}-{\gamma^2\over 8}{b_1\over c_1^2}+{2\over 3}{\beta^3\over c_1^3}
-{\beta^2\gamma\over c_1^3}+{\beta\gamma^2\over 2c_1^3}-{\gamma^3\over 12c_1^3}-$$
$$-{\beta\over 4}b_1+{\gamma\over 8}b_1-{c_1b_2\over 12}+{\beta\over 3}{b_2\over c_1}
-{\gamma\over 6}{b_2\over c_1}.\eqno(4.12)$$
and
$$c_3={3\over 16}(c_1^3-c_1^2b_1)+{c_1b_1^2\over 24}+c_1b_2-{5\over 12}\beta c_1
-{\gamma\over 4}c_1-{\beta\gamma\over 6c_1}+{\gamma^2\over 12c_1}+{\beta\over 4}b_1
+{\gamma\over 8}b_1.\eqno(4.13)$$
However, the equation (3.25) without second derivatives is identically satisfied and
gives no further relation. The same is true in $O(r^2)$, $O(r^3)$ and $O(r^4)$ (see [1]).

Assuming that this property holds in all orders,then all $a_n, n\ge 1$, $c_n, n\ge 2$ 
are determined by $c_1$ and $b_n, n\ge 1$.
That means the function $b(r)$ remains completely free. We get a class of solutions with one
arbitrary function $b(r)$. We shall return to this point in section 6.

For $r\to\infty$ we expect the solution to be asymptotically flat because it should
represent an isolated model galaxy. Therefore we set up an expansion of the form
$$a=\sum_{n=1}^\infty {A_n\over r^n}\eqno(4.14)$$ 
$$b=\sum_{n=1}^\infty {B_n\over r^n}\eqno(4.15)$$ 
$$c=\sum_{n=1}^\infty {C_n\over r^n}.\eqno(4.16)$$
Then the exponentials are expanded as follows
$$e^{b-c}=1+{B_1-C_1\over r}+{1\over r^2}\Bigl(B_2-C_2+{1\over 2}(B_1-C_1)^2\Bigl)+$$
$$+{1\over r^3}\Bigl(B_3-C_3+(B_1-C_1)(B_2-C_2)+{1\over 6}(B_1-C_1)^3\Bigl)+O(r^{-4})$$
$$e^{2b-d}=1+{B_1-A_1-2C_1\over r}+{1\over r^2}\Bigl(B_2-A_2-2C_2+{1\over 2}(B_1-A_1-2C_1)^2\Bigl)+$$
$$+{1\over r^3}\Bigl(B_3-A_3-2C_3+(B_1-A_1-2C_1)(B_2-A_2-2C_2)+$$
$$+{1\over 6}(B_1-A_1-2C_1)^3\Bigl)+O(r^{-4})$$
Substituting this into (3.24), (3.25) and (4.8) we get from $O(r^{-3})$
$$A_1=-B_1.\eqno(4.17)$$
In order $O(r^{-4})$ it follows from (3.24)
$$C_2={B_1^2\over 2}-{B_1C_1\over 2}-B_2-{C_1^2\over 4}+{\al\over 2}\eqno(4.18)$$
and (4.8) determines
$$A_2=-{B_1^2\over 2}+{B_1C_1\over 2}.\eqno(4.19)$$
From (3.25) we obtain the quadratic equation
$$0=(B_1-C_1)^2+C_1-B_1\eqno(4.20)$$
hence
$$C_1=B_1,\quad {\rm or} \quad C_1=B_1-1\eqno(4.21)$$
so that we find two possible asymptotic solutions.

In next order $O(r^{-5})$ (3.24) gives
$$C_3=-{B_1^2C_1\over 4}+{B_1B_2\over 4}+{B_1C_1^2\over 4}+{\al\over 4}B_1+{B_1\over 24}+$$
$$+{B_2C_1\over 2}-{B_3\over 2}+{C_1^3\over 8}-{\al\over 2}C_1-{C_1\over 24}\eqno(4.22)$$
and from (4.8) we get
$$A_3=-{B_1^3\over 18}+{B_1^2C_1\over 6}-{B_1^2\over 54}-{B_1B_2\over 2}-{B_1C_1^2\over 6}+$$
$$+{B_1C_1\over 27}+{\al\over 6}B_1-{B_1\over 108}-{C_1^3\over 36}-{C_1^2\over 54}
+{C_1\over 108}.\eqno(4.23)$$
Then the equation (3.25) is identically satisfied for both solutions (4.21). The same
is true in $O(r^{-6})$ and $O(r^{-7})$ (see [1]): from (3.24) one finds $C_n$, 
$A_n$ is given
by (4.8), and (3.25) gives no further restriction on the solution. As in the expansion
for small $r$, $b(r)$ is completely free and determines $a(r)$ and $c(r)$; but there is a
small difference: $c_1$ remains free for small $r$, whereas $C_1$ is fixed according to
(4.21). So if one tries to match the two expansions by numerical integration of the field
equations one better starts from large $r$ and integrates backwards to small $r$. Counting
the degrees of freedom it should then be possible to match the solution. To learn how this
can be done in a realistic case we discuss the connection of the metric with observable
quantities in the next section.

\section{Geodesics and rotation curves}

As in ordinary general relativity we assume that test particles like gas or stars in a
galaxy move along geodesics. The geodesic equation reads
$${d^2x^\al\over ds^2}+\Gamma^\al_{\beta\gamma}{dx^\beta\over ds}{dx^\gamma\over ds}=0
\eqno(5.1)$$
where $s$ is an affine parameter. We restrict ourselves to motion in the plane $\te=\pi/2$.
Using the Christoffel symbols (3.4-7) we have the following three equations
$${d^2ct\over ds^2}+a'{d\,ct\over ds}{dr\over ds}=0\eqno(5.2)$$
$${d^2r\over ds^2}+{a'\over 2}e^{a-b}\Bigl({d\,ct\over ds}\Bigl)^2+{b'\over 2}
\Bigl({dr\over ds}\Bigl)^2
-\Bigl(r+{r^2\over 2}c'\Bigl)e^{c-b}\Bigl({d\fii\over ds}\Bigl)^2=0\eqno(5.3)$$
$${d^2\fii\over ds^2}+2\Bigl({1\over r}+{c'\over 2}\Bigl){dr\over ds}{d\fii\over ds}=0.
\eqno(5.4)$$
We further restrict ourselves to circular motion $r=$const, so that $dr/ds=0$. Then from
(5.3) we get
$${a'\over 2}e^{a-b}\Bigl({d\,ct\over ds}\Bigl)^2=\Bigl(r+{r^2\over 2}c'\Bigl)e^{c-b}
\Bigl({d\fii\over ds}\Bigl)^2\eqno(5.5)$$
This implies
$$\Bigl({d\fii\over d\,ct}\Bigl)^2={a'\over 2r+r^2c'}e^{a-c}.\eqno(5.6)$$

The left-hand side in (5.6) is essentially the rotation velocity $v$
or angular velocity $\omega$
$${d\fii\over d\,ct}={\omega\over c}={v\over cr},\eqno(5.7)$$
so that we obtain the important result
$${v^2\over c^2}={a'r\over 2+c'r}e^{a-c}.\eqno(5.8)$$
We observe with satisfaction that the measurable quantity $v$ depends on the metric
functions $a(r)$ and $c(r)$, only, not on $b(r)$ which was arbitrary in the power series
expansions.

Since we have neglected ordinary matter in our model we must interpret $v(r)$ (5.8) as
the rotational velocity of the dark matter halo. Usually this halo is described by a
density profile $\ro(r)$ or mass profile
$$M(r)=4\pi\int\limits_0^r\ro(r)r^2\,d^3x.\eqno(5.9)$$
Assuming Kepler's third law
$$v^2={GM(r)\over r}\eqno(5.10)$$
the dark matter density profile is given by
$$\ro(r)={1\over 4\pi Gr^2}{d\over dr}(rv^2)={1\over 4\pi r^2}(v^2+2rvv').\eqno(5.11)$$
This profile is usually determined from numerical simulations on the basis of Newtonian
dynamics ([5] and references given therein).
On the other hand the total density (dark plus ordinary matter) can be measured by means of
the rotation curve $v(r)$ of the galaxy. We now have a third approach to this quantity with
our tensor-four-scalars theory.

Of particular interest is the behavior of $\ro(r)$ for small $r$ because it is difficult
to determine this by numerical simulation. Substituting the power series expansions (4.1)
and (4.3)
$$ra'=a_1r+2a_2r^2+3a_3r^3+\ldots$$
$$rc'=-2+c_1r+2c_2r^2+3c_3r^3+\ldots\eqno(5.12)$$
into (5.8), we get the following expansion for the circular velocity profile
$${v^2\over c^2}=r^2{a_1+2a_2r+3a_3r^2+\ldots\over c_1+2c_2r+3c_3r^2+\ldots}
{e^{a_0-c_0}\over r_c^2}[1+(a_1-c_1)r+\ldots]\eqno(5.13)$$
Since the 2 in the denominator in (5.8) is cancelled by the logarithmic terms in
(4.3), the circular velocity $v$ has a linear behavior $v(r)\sim r$ for $r\to 0$.
This can be traced back to the action of the vector graviton or scalar field in (3.19).
For the density profile (5.11) this means that {\it the tensor-four-scalars theory
predicts a flat dark density core} $\ro(r)\sim r^0$ for $r\to 0$. On the other hand the older
numerical simulations of Navarro, Frenk and White [6] have given an inner cusp
as steep as $r^{-1}$.

In this situation one hopes that observations can distinguish between the different
predictions. Probably the best galaxy for this purpose is M33. This is a low-luminosity
spiral galaxy in the Local Group which is dark-matter-dominated. A detailed study of the
radial distribution of visible and dark matter was made by E. Corbelli [7].
Corbelli has fitted the measured data using dark density profiles with both flat
and cuspy cores. Unfortunately, the fits were equally good. But in later work by Donato
et al.[8] the flat dark density core was confirmed.

\section{Reduction of the equations and discussion}

In section 4 we have found that the metric function $b(r)$ was not determined by the field
equations (3.24) (3.25) (4.8). Another way to see this is the following. We solve
equ.(3.25) for $b(r)$
$$b=c+2\log r-\log\Bigl(1-{\al\over 2r^2}e^{-a-c}\Bigl)+$$
$$+\log\Bigl({a'c'\over 2}+{a'+c'\over r}+{c'^2\over 4}+{1\over r^2}\Bigl).\eqno(6.1)$$
This enables us to eliminate $b$ in the remaining two equations (3.24) and (4.8). Then
from both equations we arrive at the same single equation between $a(r)$ and $b(r)$
$$c''={a''\over a'}\Bigl(c'+{2\over r}\Bigl)+{2\over r^2}+{2r^2a'c'+4r(a'+c')+r^2c'^2
+4\over 2r^2-\al\exp(-a-c)}.\eqno(6.2)$$
This shows the degeneracy in the system of vacuum field equations.

A similar degeneracy is known from the Schwarzschild solution with the difference that
there the two remaining equations allow to calculate the two metric functions. Including
the dark sector by means of the $v$-field, the vacuum problem without normal matter
becomes undetermined. To make it determined we need a further equation for $a$ and $c$.
For this purpose we can use equ.(5.8) for the rotation velocity in the form
$$a'e^a={v^2\over c^2r}(2+rc')e^c.\eqno(6.3)$$
Hence, if the rotation velocity is known the metric can be calculated. This implies
that {\it there is no universal dark density profile} in contradiction to earlier
numerical simulations [6]. One expects that the degeneracy in the field equations
is lifted if the ordinary matter is included in the form described in section 2.
But this is a much more difficult problem.

\end{document}